\documentclass[showpacs,preprintnumbers,english,twocolumn,amsmath,amssymb]{revtex4}
\usepackage[T1]{fontenc}
\usepackage[latin1]{inputenc}
\usepackage{graphicx}
\usepackage{amssymb}
\usepackage{verbatim}
\usepackage{epic,eepic}
\usepackage{babel}

\begin{document}

\title{Density and Temperature of  Bosons from Quantum Fluctuations}

\author{ Hua Zheng$^{a,b)}$ and Aldo Bonasera$^{a,c)}$}
\affiliation{
a)Cyclotron Institute, Texas A\&M University, College Station, TX 77843, USA;\\
b)Physics Department, Texas A\&M University, College Station, TX 77843, USA;\\
c)Laboratori Nazionali del Sud, INFN, via Santa Sofia, 62, 95123 Catania, Italy.}




\begin{abstract}
A  method to determine the density and temperature of a system is proposed based on quantum fluctuations typical of Bosons in the limit where the reached temperature T is close to the critical temperature $T_c$ for a Bose condensate
at a given density $\rho$.  Quadrupole and particle multiplicity fluctuations relations are derived in terms of $\frac{T}{T_c}$. This method is valid for weakly interacting infinite and finite Boson systems.
 As an example, we apply it  to heavy ion collisions using the Constrained Molecular Dynamics (CoMD) approach which includes the Fermi statistics.  The model shows some clusterization into deuteron and ${\alpha}$ clusters which could suggest a Bose condensate.
However, our approach demonstrates that  in the model there is no Bose condensate  but it gives useful informations to be tested experimentally.
We stress the differences with  methods based on classical approximations. The derived 'quantum' temperatures are systematically higher than the corresponding  'classical' ones.  The role of the Coulomb charge of fragments is discussed.

\end{abstract}

\pacs{ 25.70.Pq,42.50.Lc, 64.70.Tg}

\maketitle

Fragmentation of heavy ions displays an anomalous production of $\alpha$ particles as compared to nucleons \cite{1,2,10}.  This poses the question of what is the role of Bosons in nuclear matter and finite nuclei.  We know that light nuclei display a $\alpha$-cluster 
 structure which could be exemplified by the so-called 'Hoyle' state in $^{12}C$, i.e. the first excited state of such a nucleus which decays into $3\alpha$'s\cite{arnett}.  The fact that the ground state of nuclei could be made of $\alpha$
clusters could justify its copious production in heavy ion collisions near the Fermi energy.  At the same time these facts arise the natural question if $\alpha$ clustering and production could be signatures of a Bose-Einstein condensate.
\begin{figure}
\centering
\includegraphics[width=1.15\columnwidth]{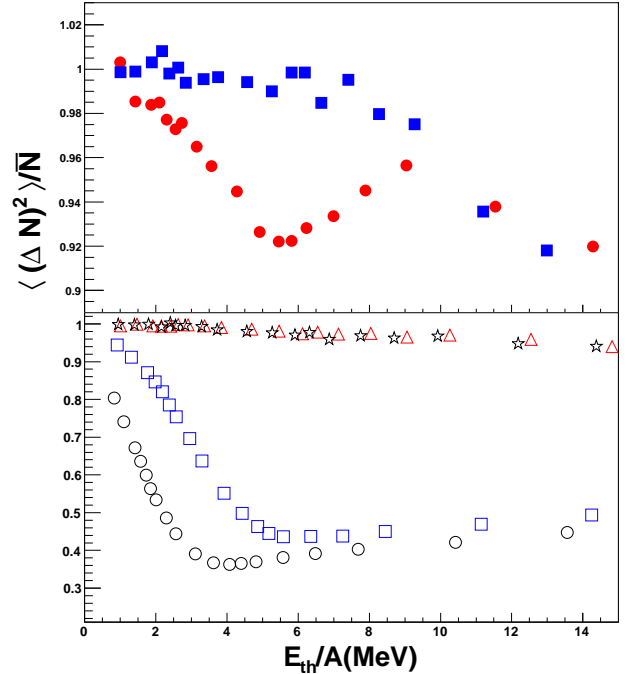} 
\caption[]{Normalized variance versus exitation energy per nucleon\cite{plb1}.
 (Top panel)  CoMD results for d (full circles)  and $\alpha$ particles (full squares).  For comparison the normalized fluctuations for fermions (bottom panel). (Open)  Circles, squares and triangles refer to protons, neutrons and tritons,
   stars refer to $^3He$. Notice the change of scales in the two panels.  }
\label{Fig1}
\end{figure}
In this work we discuss some properties at finite temperatures assuming either a classical gas or a quantum Bose system, a Fermi system has been discussed in\cite{plb1}.  We show
that at the densities and temperatures of interest the classical approximation is not valid.  This is at variance with many experimental and theoretical results in heavy ion collisions near the Fermi energy  \cite{10,albergo,pocho,15,15b,michela,15a} which assume the classical approximation to be valid. 
We base our approach on fluctuations estimated from an event by event determination of fragments arising  after the energetic collision.  A similar method has recently been applied to observe suppression of 
fluctuations in a trapped Fermi gas \cite{prl} and in Bose condensates\cite{bose}. In ref.\cite{plb1}, we proposed a method to go beyond ref.\cite{prl,bose} by including quadrupole fluctuations as well to have a direct measurement of densities and temperatures for subatomic systems for which it is difficult to obtain such informations in a direct way.
 We apply our proposal to the microscopic CoMD approach \cite{17}  which includes Fermionic statistics.  Because of antisymmetrization, the model gives some clustering into ${\alpha}$ like structure in the ground state of some nuclei such as $^{40}Ca$.  Also, in fragmentation reactions, a larger 
 yield of  ${\alpha}$ clusters is observed, however the experimental yield is largely underestimated\cite{17}.  These features should be kept in mind when discussing a possible Bose condensate in the model. In fact we do not find any clear evidence of the condensate in the model but 
 we can intuitively understand how to obtain it and especially how to pin down the condensate in experimental data.  More refined models are possible but experimental data are needed in order to guide the modeling. We believe that such data could be obtained from heavy ion collisions using 
 $4\pi$ detectors and performing a careful event by event analysis, the major serious problem we foresee is in the event selection for which the results discussed here in terms of the CoMD approach could be of guidance.  \\
A method for measuring the temperature was proposed in  \cite{15b} based on momentum fluctuations of detected particles.  A quadrupole $Q_{xy}=<p^2_x-p^2_y>$ is defined in a direction transverse to the beam axis (z-axis) to minimize non equilibium effects \cite{plb1}
and the average is performed, for a given particle type, over events. Such a quantity is zero
in the center of mass of the equilibrated emitting source.
Its  variance  is given by the simple formula:
\begin{equation}
 \sigma^2_{xy}=\int d^3p(p^2_x-p^2_y)^2f(p)
\label{1}
\end{equation}
where f(p) is the momentum distribution of particles.  In \cite{15b} a classical Maxwell-Boltzmann distribution of particles at temperature $T_{cl}$ was assumed which gives: $ \sigma^2_{xy}=\bar N 4m^2T_{cl}^2$,  m is the mass of the fragment. $\bar N$ is the average number of particles which could be conveniently normalized to one. 
 In heavy ion collisions,  the produced particles  do  $\it not$  follow
classical statistics thus the correct distribution function must be used in eq.(1).  Protons(p), neutrons(n), tritium(t)
 etc. follow the Fermi statistics while\cite{plb1}, deuterium(d), alpha($\alpha$) etc., even though they are constituted of nucleons, should follow the Bose statistics.  In this work we will concentrate on Bosons only and in particular
d and $\alpha$ which are abundantly  produced in the collisions  thus carrying important informations on the densities and temperatures reached.  
  \begin{figure}
\centering
\includegraphics[width=1.05\columnwidth]{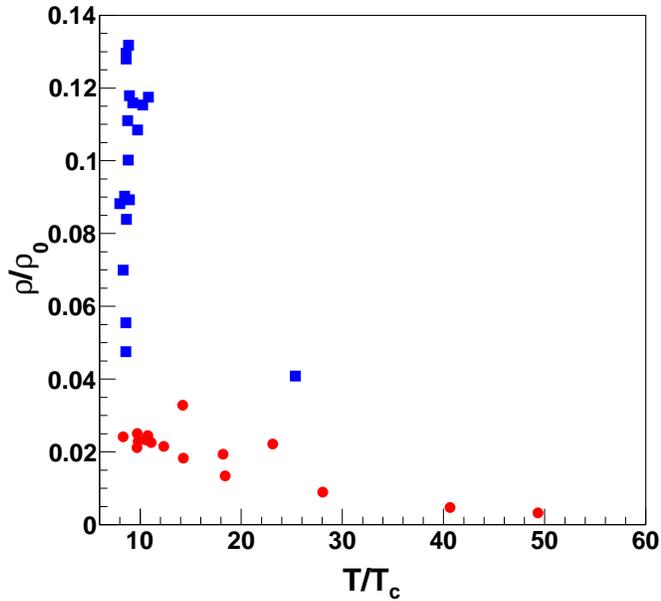}
\caption[]{Reduced density versus reduced temperature for Bosons. Symbols as in figure 1.}
\label{Fig2}
\end{figure} 
Using a Bose-Einstein distribution f(p) and expanding near the critical temperature $T_c$ at a given density $\rho$,  we get \cite{8}:
\begin{equation}
 \sigma^2_{xy}=(2mT)^2\frac{g_{7/2}(0)}{g_{3/2}(0)}	(T<T_c)
 \end{equation}
 \begin{equation}
 \sigma^2_{xy}=(2mT)^2\frac{g_{7/2}(z)}{g_{3/2}(z)}	(T>T_c)
\label{2}
\end{equation}
where the $g_n(z)$ functions are well studied in the literature\cite{8} and $z$ is the fugacity which depends on the critical temperature for Bose condensation and thus on the density of the system\cite{8}.
Notice the similarity with the classical result which is modified by the ratio of the $g_n(z)$ functions only.  Below the critical temperature such a ratio is a constant $\frac{g_{7/2}(0)}{g_{3/2}(0)}=????$, 
thus the Bose temperature is higher than the classical one.  Above the critical temperature the ratio of the $g$ functions approaches one for large T and small densities thus recovering the classical result. These features
should be contrasted with Fermion systems for which the temperature is always smaller than the classical limit\cite{plb1}, i.e. opposite to the Boson case. These results are very important and could be used to pin down a Bose condensate by comparing
Fermions and Bosons produced in nuclear reactions on an event by event basis.
The quadrupole fluctuations depend
on temperature and density through $T_c$, thus we need more informations in order to be able to determine both quantities for $T>T_c$. 

 Within the same framework we can calculate the fluctuations of the d, $\alpha$ multiplicity distributions.  These are given by \cite{8}:
\begin{equation}
 \frac{<(\Delta N)^2>}{\bar N}=(\frac{T}{T_c})^{3/2}[1+(\frac{T}{T_c})^{3/2}]  (T<T_c)
\label{3}
\end{equation}
\begin{equation}
 \frac{<(\Delta N)^2>}{\bar N}=0.921\frac{(\frac{T}{T_c})^{3}}{[1-(\frac{T}{T_c})^{3/2}]^2}  (T>T_c)
\label{33}
\end{equation}
The difference with the Fermionic case\cite{plb1} is striking: for Bosons, fluctuations are larger than the average and might diverge near the critical point, eq.(5),
 in the indicated approximations.  Interactions and finite size effects will of course smoothen 
the divergence\cite{8}.
 Two solutions are possible depending if we are above or below the critical temperature for a Bose condensate.  Below the critical point, eq.(2) can be used to calculate T and then eq.(4) gives the critical temperature and the corresponding
density.  Above the critical point it is better to estimate the chemical potential which, in the same approximation, is given by:
\begin{equation}
 \frac{{-\mu}}{T}= \frac{1}{2}\frac{1}{\frac{<(\Delta N)^2>}{\bar N}} (T>T_c)
\label{37}
\end{equation}
Notice the similarity of this result to eq.(3) in \cite{plb1}, where the chemical potential is substituted by the Fermi energy.
From this equation we can estimate the $g_n$ functions entering eq.(3) and obtain the value of T.  Using such a value in eq.(5), gives $T_c$ and the density $\rho$.  In the numerical simulations discussed below we can always
use the two solutions but one of these can be rejected using physical considerations.  For instance assuming that we are below the critical point leads to densities as high as ten times the ground state density which is unphysical 
for heavy ion collisions around the Fermi energy.

  To illustrate the strength of our approach we simulated
$^{40}Ca+^{40}Ca$ heavy ion collisions at fixed impact parameter $b=1fm$ and beam energies  $E_{lab}/A$  ranging from 4 MeV/A up to 100 MeV/A.  Collisions were followed up to a maximum time 
$t=1000 fm/c$ in order to accumulate enough statistics. Because of the low yield of fragments, more than 200,000 events per beam energy were calculated. 
In order to correct for collective effects as much as possible, we defined a 'thermal' energy as in ref.\cite{plb1}.

 \begin{figure}
\centering
\includegraphics[width=1.05\columnwidth]{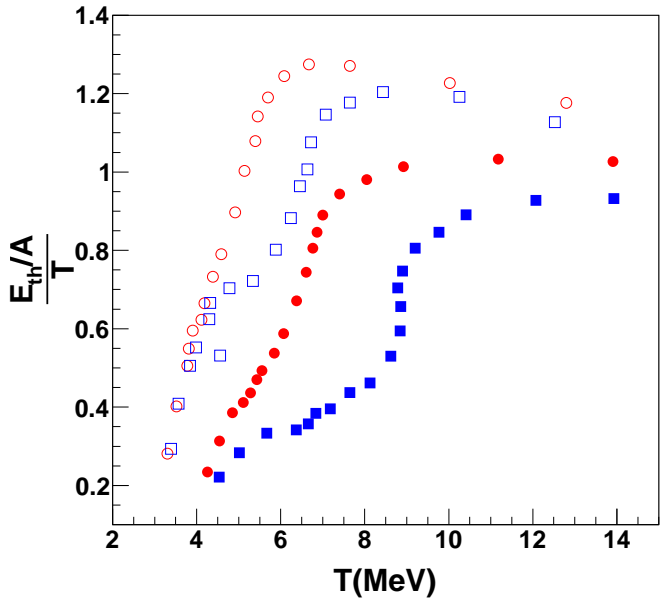} 
\caption[]{Excitation energy per particle divided by temperature vs temperature. Symbols as in figure 1, open symbols refer to calculations without Coulomb.}
\label{Fig3}
\end{figure}  
In figure 1 we plot the reduced variances versus excitation energy per particle\cite{plb1}.
   The CoMD results are given by the full symbols, top panel.
     As we see in the figure, $\alpha$ normalized fluctuations are generally larger than d-fluctuations.  As we will show below, this implies
that those particles might explore different regions of densities and temperatures.  
In both cases, fluctuations are large and,  in some cases, above  Poissonian for  $\alpha$'s.  In order to understand if a Bose condensate occurs in the model (and in the future in experiments) it is 
instructive to compare the Bosons normalized fluctuations to those of Fermions discussed in ref.\cite{plb1}. In figure 1-bottom panel, normalized Fermions fluctuations are given. 
 As we see the normalized fluctuations of p and n are much smaller than 1
at variance with the Bosons case, which would suggest a condensate.  However, heavier Fermion clusters such as $^3He$ and tritons, display fluctuations larger than d and smaller than $\alpha$.  
In particular $^3He$ fluctuations are smaller than t which demonstrates that it is not 'directly' a Coulomb effect.
These facts are important to understand what is happening in the model and eventually search for an experimental confirmation. We offer here an intuitive explanation of the relative role of normalized fluctuations for different particles type.  The CoMD model is essentially classical
with a constraint in the equation of motion which keeps the occupation probability $\bar f(r,p,t)$ smaller than 1 as dictated by the Pauli principle for Fermions\cite{1,2,17}. A further implementation of the Pauli principle is in the collision term which avoids that colliding nucleons
occupy phase space regions which are occupied by other nucleons.  Thus the Pauli principle reduces the available phase space and in turn the normalized fluctuations.   For this reason p and n fluctuations are smaller than Poissonian.  When composite fragments are formed, d,t, etc..,
the effect of Pauli blocking is reduced (also because those particles form at low densities, see below), thus fluctuations become comparable to their average value.The effect that reduces the available phase space is now the binding energy.  Not all nucleons can form a bound state,
especially if their relative kinetic energies are larger than the potential one.  For this reason d fluctuations are smaller than $^3He$ which are smaller than t and smaller than  $\alpha$ fluctuations.  In particular, since the d is over bound in the model (about 7 MeV) we expect the data to exhibit 
smaller fluctuations.  On the other hand the binding energy of  $\alpha$ is smaller than the data (about 20MeV) thus suggesting the possibility of larger fluctuations than displayed in figure 1.  However,  the important ingredient which is missing in the model is the possibility of boson-boson
collisions ( $\alpha$- $\alpha$, d-d, etc.) and correlations.  If such collisions occur their probabilities will be enhanced by the Bose factors $(1+\bar f(r,p,t))$ in contrast to the Pauli blocking factors\cite{1,2}.  This will produce fluctuations larger than Poissonian, which is a signature of a Bose condensate.
Of course the last word is in the experimental data, similar to trapped Fermi and Boson gases\cite{prl,bose}. Notice in figure 1 the occurrence of a minimum at a similar excitation energy for p,n and d but not for heavier clusters.
This will have an effect on the EOS as we will show below.\\
 \begin{figure}
\centering
\includegraphics[width=1.15\columnwidth]{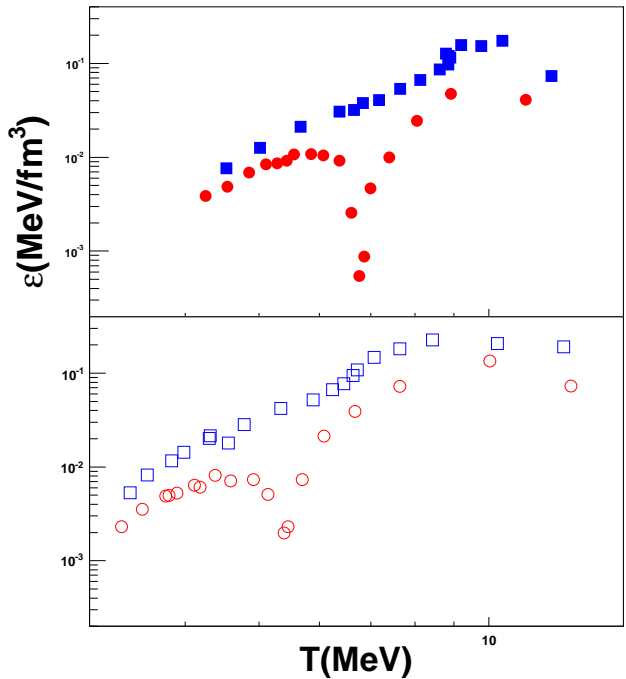} 
\caption[]{Energy density versus temperature, top panel with Coulomb, bottom panel without Coulomb. Symbols as in figure 3.}
\label{Fig4}
\end{figure}
It is interesting to discuss the densities 'seen' by the different Bosons during the reaction.  A plot of density (divided by the ground state density) versus temperature (divided by the
critical temperature for a condensate) is given in figure 2. Notice the peculiar behavior of d and $\alpha$ clusters.  While the latter are formed at a constant reduced temperature but at different densities for each
beam energy, the deuterons are formed always at a very small (constant) density but at different temperatures.  As we noticed above two effects are at play.  The first is that there is no Pauli blocking
for nucleons inside the clusters, the second is the different binding energies.  Since d are over bound in the   model we expect to see that even smaller densities will be seen in the data, the opposite we
expect for $\alpha$.  These features remind of Mott transitions and in particular suggest that different particle types might be sensitive to different regions of the NEOS as already noticed for Fermion\cite{plb1}.
This is an interesting results since it gives the opportunity to determine the EOS even though we are dealing with small, dynamical systems, where we need to correct for some effects such as Coulomb.\\
In figure 3 we plot the thermal energy divided by T versus temperature.  From eq.(2-3) we expect to get the classical result at very high T and this is the reason for the y-axis choice in the figure.  As we see the results
indeed approach the value of $3/2$ as expected, even though we notice a decrease at the largest temperature which is an indication of transparency in the model.
  For some calculations we have also turned off the Coulomb field (open symbols).  It is evident that the general features are the same apart from a shift in T, similar to what
was observed for p and n\cite{plb1}.  It is interesting to notice that $\alpha$ particles indicate a first order phase transition, while d might suggest a crossover or a second order phase transition.  Recall that the densities explored
by the two fragments are quite different, thus it is not surprising if they behave differently.  These might be better grasped in figure 4 where the energy density is plotted versus temperature.
This figure should be compared to the Fermion case\cite{plb1} where a rapid increase is noticed at about 3-4 MeV temperature.  Such an increase is seen in the d-case but there is a deep minimum near such temperatures,
reflecting the fluctuations displayed in figure 1.  As we see in figure 4 (bottom panel), such a behavior is not due to Coulomb which simply shifts the temperatures to higher values.
Of course a mixture of bosons and fermions produces a quite rich EOS as in the case of $^3He$, $^4He$ gas mixtures\cite{8}, thus our proposed approach might open a quite interesting field of research.\\
In conclusion, in this work we have addressed a general approach for deriving densities and temperatures of bosons.   
In the framework of the  Constrained Molecular Dynamics model, 
we have discussed collisions of heavy ions
below 100MeV/A and obtained densities and temperatures at each bombarding energy. Knowing the thermal energy of the system, we can derive the energy density and temperature reached during the collision.  
  We have seen 
in this work that different particles like d and ${\alpha}$ explore different density and temperature regions, similar to the Fermion case\cite{plb1}. Open problems such as a Bose condensate in nuclei,
Mott transitions, pairing etc. in low density matter 
might be addressed through a detailed study of the EOS. 

\end{document}